\documentclass[fleqn,usenatbib]{mnras}

\usepackage{graphicx}	
\usepackage{amsmath}	
\usepackage{amssymb}	
\usepackage{tablefootnote}

\usepackage[export]{adjustbox} 

\usepackage{color}

\definecolor{revisiongreen}{RGB}{0,95,0}
\DeclareRobustCommand{\rev}[1]{#1}

\usepackage{newtxtext,newtxmath}

\usepackage[T1]{fontenc}

\DeclareRobustCommand{\VAN}[3]{#2}
\let\VANthebibliography\thebibliography
\def\thebibliography{\DeclareRobustCommand{\VAN}[3]{##3}\VANthebibliography}

\title[New JCMT-BISTRO results of M16]{The magnetic field in M16: new results from the JCMT BISTRO survey}
\author[R.M.R. Tulloh et al.]{R.M.R. Tulloh$^{1}$\thanks{E-mail: rmrtulloh@lancashire.ac.uk}, D. Ward-Thompson$^{1}$, 
J. Karoly$^{1,2}$, J.M. Kirk$^{1}$, K. Pattle$^{2}$\\
\\
$^{1}$Jeremiah Horrocks Institute, University of  Lancashire, Preston, PR1 2HE, UK\\
$^{2}$ Department of Physics and Astronomy, University College London, Gower Street, London, WC1E 6BT, UK\\
\\
Accepted 2026 xxx xx. Received 2026 xxx xx; in original form 2026 xxx xx. }

\begin{document}

\date{}

\pagerange{\pageref{firstpage}--\pageref{lastpage}} \pubyear{2026}

\maketitle

\label{firstpage}

\begin{abstract}   
We present a study of the magnetic field (B-field) in the 
Eagle Nebula (M16) using data from the
James Clerk Maxwell Telescope (JCMT) B-fields In 
STar-forming Region Observations
(BISTRO) survey. We extend our previous study of 
the well-known `Pillars of Creation'
region of the nebula to study two other structures in the region, 
the `Spire' and the bright-rimmed cloud (BRC) known as `SFO30'.
These structures show similar B-field morphologies to the Pillars. 
We calculate B-field strengths using the 
Davis-Chandrasekhar-Fermi (DCF) method and thereby 
estimate the magnetic potential energy in each of these regions
for comparison with
the local gravitational, turbulent and thermal energies, 
as well as the pressure energy of the shocks from nearby O-type stars.
We find approximate
equipartition between the energies supporting the regions and those
tending towards collapse, 
with the B-field support playing a significant role. This
agrees with
our earlier hypothesis that B-fields help support
pillar-like structures in molecular clouds and make them longer-lived than they would otherwise have been.
\\
\end{abstract}


\section{Introduction}

The Hubble Space Telescope (HST) has taken many images, one of the most famous being the “Pillars of Creation” in the Eagle Nebula. These photo-ionized pillars, or columns, are star-forming regions associated with nearby intermediate- to high-mass stars ($\gtrsim~\!8\,M_\odot$). \citet{2018ApJ...860L...6P} made the first detailed measurements of the magnetic ﬁeld (B-field) in the densest parts of the Pillars, as part of the B-fields in STar-forming Region Observations (BISTRO) survey using the James Clerk Maxwell Telescope (JCMT) in Hawaii \citep{2017ApJ...842...66W}. The present study aims to develop this work further.

\cite{1949Sci...109..165H}, and \cite{1949AJ.....54..187H} showed that the light which arose from areas with significant extinction was plane polarized. \citet{1951ApJ...114..206D} proposed that it was dust grains that were responsible for this effect, but the mechanism was unclear and still remains an area of controversy. The Radiative Alignment Torque theory \citep{1976Ap&SS..43..291D}  proposed that anisotropic radiation fields exert torques on rotating asymmetric dust grains, leading to partial alignment of their angular momentum vectors. The grains precess about the ambient B-field direction. These grains 
are paramagnetic (containing silicates or iron) and above the critical size ($10^{-6}\,\mathrm{m}$), below which, rotation is primarily thermal \citep{2007JQSRT.106..225L}.  The re-emission in the far infra-red spectrum results in 
partial linear polarization which is perpendicular to the B-field \citep{2008ApJ...676L..25L}.  

Young massive stars produce high-energy photons which ionize the surrounding parent molecular cloud, providing a shockwave which erodes the cloud \citep{2007ARA&A..45..481Z}. Complex structures can form in the dense gas at the bow shock interfaces \citep{1954ApJ...120....1S} and particularly dense neutral columns, or pillars, can be seen protruding into photo-ionized regions, as in M16 \citep{2018ApJ...860L...6P}.

\begin{figure*}
    \centering
    \includegraphics[width=1.0 \linewidth]{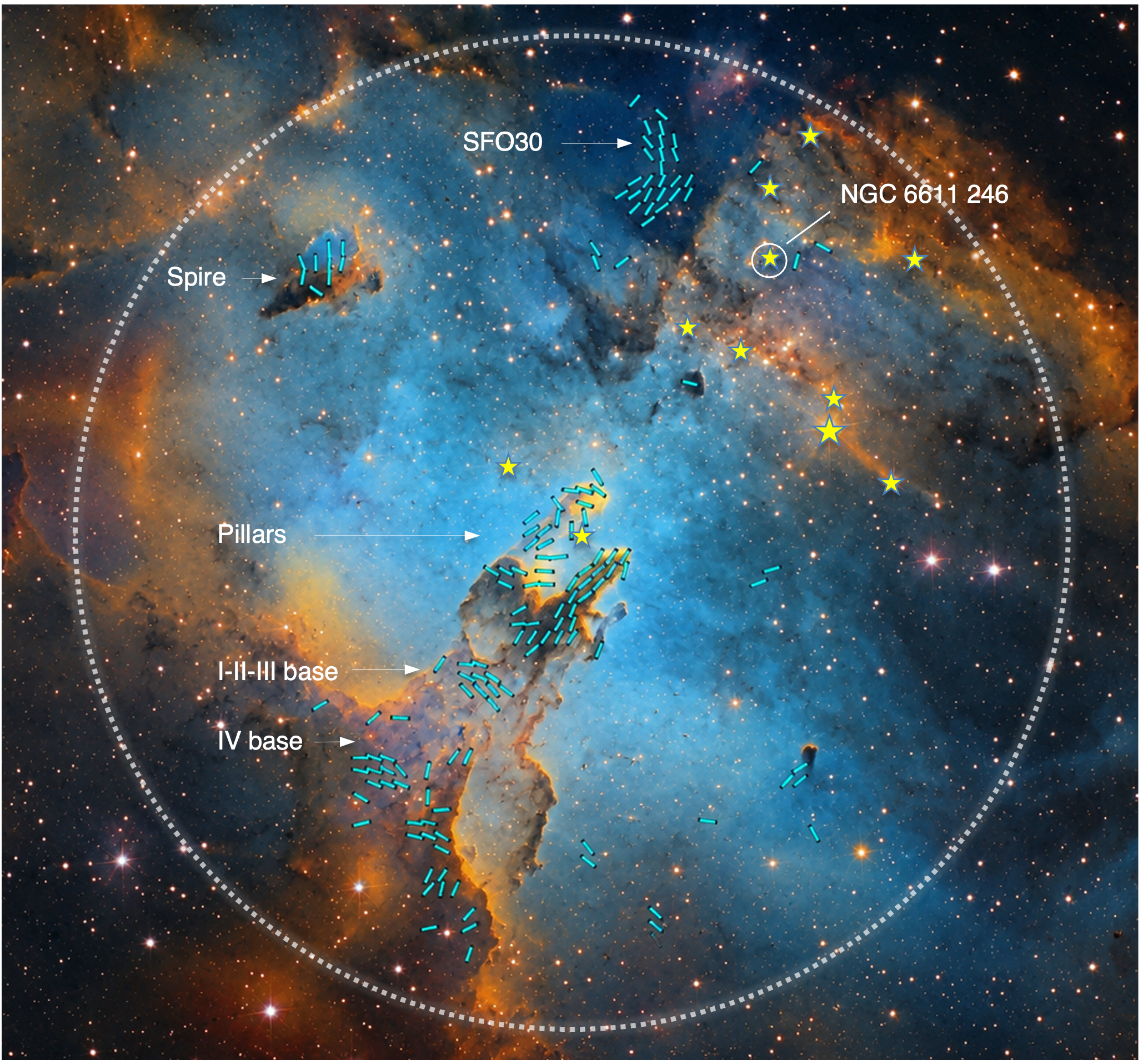}
    \caption{
Wide-field view of the Eagle Nebula (M16), based on a recoloured DSS2 Blue image obtained from photographic plates from the UK Schmidt Telescope. The four regions of study: Spire, SFO30, I-II-III base and IV base are labelled. Cyan segments show the inferred plane-of-sky magnetic-field orientation derived from 850-$\mu$m dust polarization observations from the BISTRO survey, carried out with the JCMT. The dashed boundary indicates the approximate extent of the BISTRO mapped region. Yellow stars mark selected massive members of the ionizing cluster NGC\,6611. The O7\,II star NGC 6611 246 (BD$-13^{\circ}4927$) is highlighted; its projected proximity to both the Spire and SFO30 suggests that it may contribute significantly to the local ionization and feedback environment within these structures.
}
    \label{fig1}
     \par\smallskip
    \noindent\textbf{Alt text:} Wide-field optical image of the Eagle Nebula (M16), showing the Pillars of Creation, Spire, SFO30, I-II-III base and IV base. Cyan segments show the inferred plane-of-sky magnetic-field orientations from JCMT BISTRO observations. Yellow stars mark selected massive members of NGC 6611, and a dashed circle indicates the approximate BISTRO mapped region.

\end{figure*}

Submillimetre continuum polarization surveys represent a powerful technique for detecting B-field orientation and trying to understand its role in the star formation process within molecular clouds \citep{2017ApJ...842...66W}. It is unclear whether the magnetic field stabilizes the dust and gas, or how it directs the accretion of the interstellar medium into the cores forming protostars, or what its role is in reducing star forming efficiency \citep{1999A&A...342..233W, 2001MNRAS.327..788W, 2023ASPC..534..193P}, but these are some of the questions that the BISTRO survey aims to address.

Figure 1 shows the M16 region, known as the Eagle Nebula. On it are marked the positions of the `Pillars', the young cluster of O- and B-type stars known as NGC 6611, the area mapped with SCUBA-2/POL-2 (white dashed circle) and the regions known as the `Spire' and `SFO30' -- number 30 in the catalogue of Bright-Rimmed Clouds (BRC) compiled by \citet{1991ApJS...77...59S}, which appears to be not too dissimilar in shape and perhaps formation mechanism, to the `Pillars'. Also labelled are two other regions that we looked at, known as `I-II-III base' and `IV base'.
We have previously studied how the shock fronts from the nearby young high-mass cluster NGC 6611 
photoionize and interact with the heads of the `Pillars of Creation' \citep{2018ApJ...860L...6P}. 

This paper presents the results from the BISTRO survey of the wider M16 field, shown in Figure~1, to examine other regions of the Eagle Nebula. We study the B-field deduced from the data and attempt to ascertain the role played by the B-field in this wider region.

\section{Observations}\label{sec:obs}

We surveyed the Eagle Nebula, M16, at 850 $\mu$m with SCUBA-2 \citep{2013MNRAS.430.2513H} and POL-2 \citep{2016SPIE.9914E..03F} between 6 June 2017 and 27 July 2017, under JCMT project code M17BL011.

The region was observed 20 times for $\sim 40$ minutes each, giving a total on-source integration time of $\sim 14$ hours. The observations were made in the POL-2 DAISY mode, which produces a map with high signal-to-noise ratio (SNR) in the central 3-arcmin-diameter region with increasing noise towards the edges \citep{2016SPIE.9914E..03F}. 

During the observations, which were taken in Band 2 weather, the atmospheric opacity, $\tau$ at 225\,GHz, varied between $\sim 0.05$ and $\sim 0.08$ \citep{2018ApJ...860L...6P}. The effective beam size of JCMT 
at 850~$\mu$m is $14.1$~arcsec, or $\sim 0.1$\,pc at the distance of M16, 
which we take to be 1.74 ± 0.13 kpc \citep{2019ApJ...870...32K}.    

\begin{figure*}
    \centering
    \includegraphics[width=1.0 \linewidth]{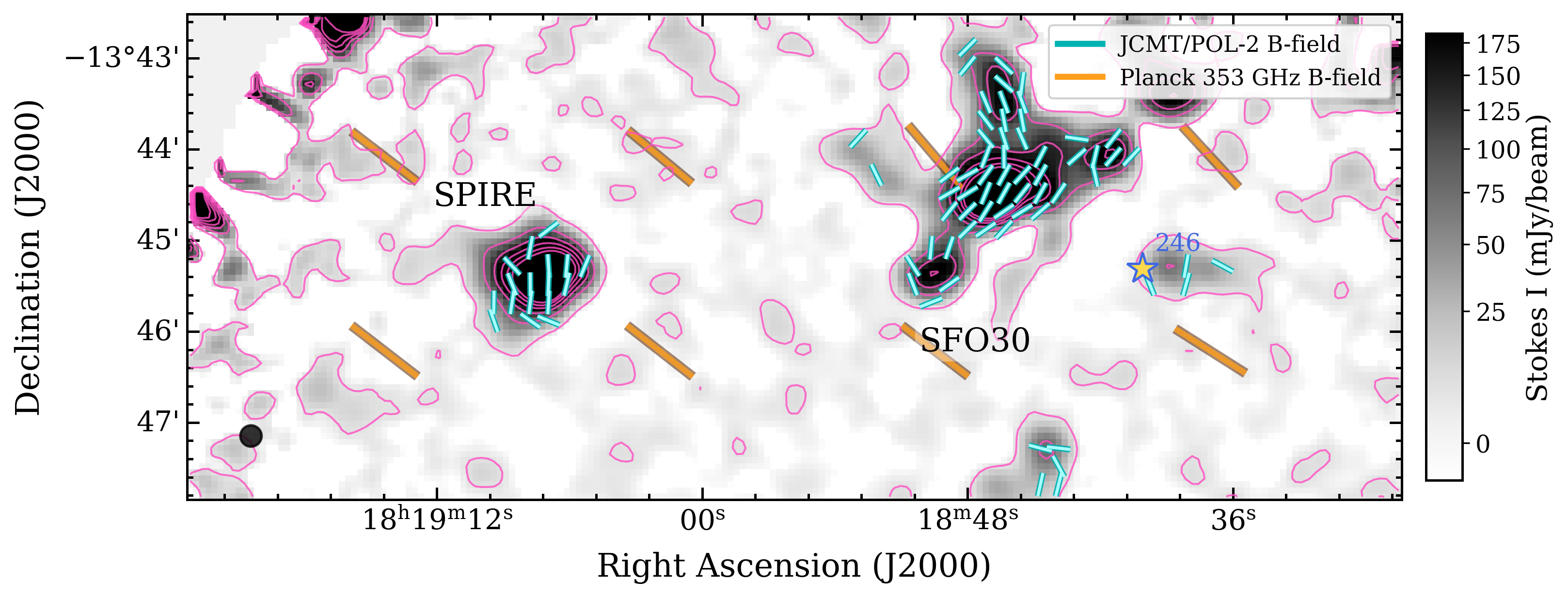}
    \caption{\rev{SCUBA-2 image of the Spire and SFO30 regions of the Eagle Nebula, M16. An ionizing star (NGC 6611 246) is also shown. The grey scale shows the smoothed Stokes I 850-$\mu$m emission, while magenta contours mark increasing intensity levels and cyan bars show the inferred plane-of-sky B-field orientation. The longer, sparse orange bars are a visual guide to the large-scale magnetic-field orientation inferred from Planck 353-GHz Stokes Q and U data. The mean Planck magnetic-field position angle within a 30-arcmin aperture centred on M16 is $44.9^{\circ} \pm 1.6^{\circ}$, with a spatial angular dispersion of $15.3^{\circ}$. The mean JCMT field orientations are $1^{\circ} \pm ±9^{\circ}$ in the Spire and $138^{\circ} \pm ±5^{\circ}$ in SFO30, corresponding to axial offsets of $43.9^{\circ} \pm ±9.1^{\circ}$ and $86.9^{\circ} \pm ±5.3^{\circ}$ respectively.
    }}
    \label{fig2}
    \noindent\textbf{Alt text:} Greyscale 850-micrometre Stokes I map of the Spire and SFO30 regions of M16, with magenta intensity contours. Cyan segments show the JCMT/POL-2 plane-of-sky magnetic-field orientations and orange segments show the large-scale Planck 353-GHz magnetic-field orientation. NGC 6611 246 is marked to the right of SFO30.
\end{figure*}

The 850-$\mu$m data were reduced using the \textsc{pol2map} routine
in the \textsc{smurf} package in \textsc{starlink}.
The standard reduction process is described in detail by \citet{2021ApJ...907...88P}. We adopted the ``August 2019'' instrumental polarization (IP) model in order to correct for the IP in the $Q$ and $U$ maps. 
Table 1 lists the coordinates of the four main regions of study. 

The final Stokes $I$, $Q$, $U$ maps and the polarization catalogue are gridded to a default 4~arcsec pixel$^{-1}$ scale. For the analysis, we follow \cite{2018ApJ...860L...6P},
who used the {\it kappa} ``\texttt{sqorst}'' command to rebin the $I$, $Q$, and $U$ maps to a 14.1~arcsec pixel$^{-1}$ scale (to match the beam size). 

Independent polarization half-vectors are measured on a 14.1~arcsec pixel$^{-1}$ scale, but the maps we show are of the B-field half-vectors, which are rotated by 90$^\circ$ relative to the polarization, while the Stokes $I$ maps we use are at the default 4~arcsec pixel$^{-1}$ scale. The survey observing strategy, data acquisition, reduction and absolute calibration are discussed in more detail by \cite{2017ApJ...842...66W}.

\section{Data Analysis}

\subsection{Magnetic field morphologies}

Figure 1 also shows overlaid a map of the B-field in those regions where we have sufficient signal to noise ratio. The half-vectors show the direction of the B-field, since they are rotated by 90$^\circ$ relative to the measured polarization vectors. We mark the four additional regions we studied, that we label accordingly. The two regions at the base of the pillars are labelled `I-II-III base' and `IV base'. The other two are the `Spire' and `SFO30' regions mentioned above. Figure~2 shows a close-up of the B-field maps of the latter two regions.

Figure 1 reveals a range of B-field morphologies in the four mapped areas of molecular cloud. \cite{2018ApJ...860L...6P} suggested that the magnetic field vectors in `Pillar 2' are at right angles to a bow-shock that could be arising from the effects of the O-type star NGC 6611 205. 
We find the same vector orientation, but note that the O-type star NGC 6611 367 could also be the source of the bow-shock creating `Pillar 2'.

Figure 5 of \cite{2018ApJ...860L...6P} illustrates the proposed hypothesis that the B-field delays the erosion of the molecular cloud when an over-density is encountered. The B-field is compressed and ordered by the presence of the shock front and supports the forming pillar against lateral erosion as seen in panel (c) of that figure.

In the region that we label `I-II-III base', the field appears to be 
roughly as in the initial conditions in the Pattle model, since it is being shielded by the pillars. There is some evidence of the field turning at the edge, where it is less shielded.

The morphology of the B-field in the region we label `IV base' is somewhat more complex. This region actually appears to be slightly foreground to the pillars. There is a region that is elongated parallel to the pillars, where the field appears to be parallel to the elongation, that might be a smaller version of `Pillar 2'. At the base of this, there is a region of B-field roughly parallel to that in the `I-II-III-base' region. This then curves round to follow the edge of the cloud.

\rev{Figure 2 shows a close-up view of the regions we call the Spire and SFO30.
Once again we plot the magnetic field vectors that we have measured.}
The `Spire' appears to be similar to a pillar, with B-field vectors at its tip being parallel to a putative bow-shock. This could possibly be arising from the influence of the nearest O-type star, NGC 6611 246, which lies at $\sim$~20~pc from M16 (based on its Gaia distance). The B-field vectors in the region we label `SFO30' could also be interpreted as lying parallel to a putative ionizing front, also from NGC 6611 246.

\rev{Figure 2 also shows the large-scale magnetic field, as measured by Planck. In the 
case of the Spire, we see that the small-scale field has rotated by roughly 45$^\circ$ 
relative to the large-scale field, and for SFO30 it has rotated by roughly 90$^\circ$.
So the small-scale field has become dissociated from the large-scale field. This is
consistent with the findings of \cite{2020ApJ...899...28D}, who found that there was a critical length scale, typically between the resolution of Planck and SCUBA-2, at which the magnetic field became structured and the small-scale field dissociated from the large-scale field.}

\subsection{Masses of regions of molecular cloud}

The total (gas + dust) mass ($M_{\mathrm{T}}$) can be estimated from the dust mass by measuring the submillimetre flux density, \citep{2011isf..book.....W}, which traces the dust emission, and assuming a canonical dust-to-gas mass ratio of $\sim$~100, using equation~1:

\begin{equation}
M_{\mathrm{T}} = \, \frac{S_\nu\,D^2}{\kappa_{\nu(850)}B_\nu(T)} \, \, 
\end{equation}

\noindent
where $S_\nu$ is the integrated 850-$\mu$m flux density, D is the distance to M16, which we take to be 1.74 ± 0.13 kpc \citep{2019ApJ...870...32K}, and temperature T = 20 K \citep{1999A&A...342..233W,2021ApJ...907...88P,2023AJ....165..198H}. $\kappa_\nu$ is the gas-plus-dust mass opacity, which we take to be 0.00125 m$^2$ kg$^{-1}$ \citep{2021ApJ...907...88P}, and $B_\nu(T)$ is the Planck function at $T=20$~K. The total 850-$\mu$m flux density was integrated over an aperture of twice the FWHM in each region to find the total mass, M$_T$. 

We measured the flux density, and re-calculated the mass, of `Pillar~2'. We found it to be 35 $\pm$ 5 M$_\odot$, similar to that published by \cite{2001MNRAS.327..788W}, who obtained a value of 31 M$_\odot$, within the uncertainties, giving us some confidence in our calibration. We went on to measure the other regions and calculate their masses, and we list these in Table~1.

\begin{table}
\caption{
Physical parameters of four regions in M16. Coordinates are centroid positions. $B_{\rm pos}$ is the derived magnetic field strength, using the compact DCF relation. $\sigma_{\theta}$ is the standard deviation of the polarization vector angles. 
}
\renewcommand{\arraystretch}{1.2}\
\centering
\begin{tabular}{lcccc}
  \hline
Region & $\alpha_{\rm J2000}$ & Total Mass  & $\sigma_\theta$ 
& $B_{\rm pos}$  \\
& $\delta_{\rm J2000}$ &($M_T/M_\odot$) & $^{\circ}$ & ($\mu\mathrm{G}$)\\
\hline
Spire           
& 18{:}19{:}08.011  & 30 $\pm$ 5 & 24 $\pm$ 5 &
 $\sim$ 130 $-$ 200  \\
& $-13{:}45{:}20.63$&&&
\\
SFO 30           
& 18{:}18{:}44.090 & 38 $\pm$ 5 & 14 $\pm$ 5 &
 $\sim$ 260 $-$ 350  \\
& $-13{:}44{:}07.97$ &&& 
\\
I--II--III base 
& 18{:}18{:}58.728  & 15 $\pm$ 5
& 12 $\pm$ 5 &
 $\sim$ 130 $-$ 200 \\
& $-13{:}51{:}41.50$&&&
\\
IV base         
& 18{:}19{:}01.024 & 10 $\pm$ 5
& 18 $\pm$ 5 &
 $\sim$ 30 $-$ 50   \\
& $-13{:}54{:}19.03$ &&&
\\
\hline
\end{tabular}
\end{table}

\subsection{Magnetic Field Strengths}

The Davis-Chandrasekhar-Fermi, or DCF, method \citep{1951PhRv...81..890D,1953ApJ...118..113C}
takes an estimate of the ratio of the non-thermal velocity dispersion to the
dispersion in the magnetic field direction as a proxy for the magnetic field strength.
The two dispersions can be measured, and this
provides an estimate of the plane-of-sky B-field strength, $B_{\rm pos}$ \citep{2018ApJ...860L...6P}. In full, this is given by: 

 \begin{equation}
   B_{\text{pos}} = Q \sqrt{(4\pi \rho)} \frac{\sigma_v}{\sigma_\theta} \approx 9.3 \sqrt{\text{n(H}_2)} \frac{\Delta v}{\sigma_\theta} \mu\mathrm{G} \, ,
 \end{equation}

\noindent
where the terms have the following meanings: Q is a factor that accounts for variation in the field on scales smaller than the beam and we take a value of 0.5 $\pm$ 0.1 \citep{2001ApJ...546..980O,2012ARA&A..50...29C}; $\rho$ is the gas density of each dense core, which we have converted to hydrogen number density, $n(\mathrm{H}_2)$, assuming the core to be spherical, and that $\mu$, which is the mean molecular weight, is 2.8 \citep{2004ApJ...600..279C,2008A&A...487..993K}; $m_{\mathrm{H}}$ is the mass of the hydrogen atom.
$\sigma_v$ is the non-thermal gas velocity dispersion and $\Delta$v is the full width calculated at half maximum (FWHM) of the velocity, assuming a Gaussian velocity distribution, for which we take a value of 0.8 $-$ 1.2 km~s$^{-1}$ \citep{1999A&A...342..233W, 2018ApJ...860L...6P, Friesen_2017}; $\sigma_\theta$ is the standard deviation of the polarization angle about the mean field direction,  listed in Table~1. Using equation 2, we calculate $B_{\rm pos}$ in each of the four regions, and list these in Table 1. Note that there are large uncertainties with this method \citep{2022MNRAS.514.1575C}. \rev{
A limitation of the DCF analysis is that the angular dispersion measured from the POL-2 vectors samples the plane-of-sky structure, whereas the velocity dispersion is measured along the line of sight. The DCF estimate therefore assumes that the gas contributing to the adopted linewidth samples a depth comparable to the region over which the polarization angles are measured; if unrelated material along the line of sight contributes significantly to the linewidth, the derived magnetic-field strength may be overestimated. We believe the latter to be unlikely, but for this reason, we treat the DCF field strengths as approximate, order-of-magnitude estimates rather than precise measurements.
}

\section{Energy balance}

\subsection{Criticality}

The virial theorem equates the different forms of potential energy in a system and requires them to balance for the system to be in equilibrium. In this case, the relevant energies are the thermal dynamical potential energy (\textbf{$E_{\mathrm{T}}$}), the non-thermal (or turbulent) dynamical potential energy (\textbf{$E_{\mathrm{NT}}$}), the magnetic potential energy (\textbf{$E_{\mathrm{B}}$}), the gravitational potential energy (\textbf{$E_{\mathrm{G}}$}) and the potential energy of the ionizing radiation pressure (\textbf{$E_{\mathrm{X}}$}) from the nearby O-type stars \citep{2006MNRAS.369.1201W}. 
The first three act to support a core, whereas the last two act to make a core collapse, and so the relation between them all tells us about the stability of a core. This is known as magnetic criticality.
Consequently, for equilibrium we require that:

\begin{equation}
\begin{gathered}
\textbf{$2E_{\mathrm{T}}$} + \textbf{$2E_{\mathrm{NT}}$} + \textbf{$E_{\mathrm{B}}$}  \approx 
\textbf{$E_{\mathrm{G}}$} + \textbf{$E_{\mathrm{X}}$}  \,  
\end{gathered}
\end{equation}

\noindent
If the right-hand side is significantly greater than the left, 
we say that the core is
super-critical, and we predict that the core will collapse.
If it is significantly less than the left-hand side, 
then we say that the core is
sub-critical, and we predict that the core will expand.
The equality leads to approximate stability and equilibrium.
The terms in this equation are given by

\begin{equation} 
\textbf{$2E_{\mathrm{T}}$} = \frac{3Mk_B T}{\mu m_{\mathrm{H}}}
\end{equation}

\begin{equation} 
\textbf{$2E_{\mathrm{NT}}$} =  \frac{3M \Delta v^2}{8 \ln(2)} 
\end{equation}

\begin{equation} 
\textbf{$E_{\mathrm{B}}$} =  \frac{B^2 V}{2 \mu_0} 
\end{equation}

\begin{equation} 
\textbf{$E_{\mathrm{G}}$}= \frac{\eta  G M^2}{R} 
\end{equation}

\begin{equation} 
\textbf{$E_{\mathrm{X}}$} =  \frac{4 R^2  k_B T_{\mathrm{II}}}{D} \times 
\left(\frac{3\pi NR}{\alpha_* }\right)^{1/2} 
\end{equation}

\noindent
\citep{2006MNRAS.369.1201W},
where symbols have the following meanings,
$R$ is the core radius, $T_{\mathrm{II}}$  is the canonical temperature of the ionized H\,{\sc ii} region (we take 10$^4$ K), N is the photon rate of the ionizing star. 

We calculate that the nearest ionizing star to both the `Spire' and `SFO30' is NGC 6611 246 (aka BD-13$^\circ$4927), which, according to its Gaia DR3 distance, is $\sim$~20~pc from both \citep{2016A&A...595A...1G}.
It is an O7 II star, so we can estimate N to be  $\sim$ 3 $\times$ 10$^{47}$ to 3 $\times$ 10$^{48}$  s$^{-1}$ \citep{1997A&A...322..598S, Telford_2023}. $\alpha_*$ is the recombination coefficient for atomic hydrogen into excited states only ($2\times10^{-19}\,\mathrm{m^3\,s^{-1}}$). $\eta$ is the coefficient determined by the density profile of the core, where we take $\eta$ = 0.6 $-$ 0.8 to allow for different density profiles \citep{2001ApJ...547..317W,2006MNRAS.369.1201W}. In the calculation of \textbf{$2E_{\mathrm{T}}$} we allow for a range of values of $T$ of up to $\sim$~30~K.

Following previous analyses in BISTRO publications, we adopt the one-dimensional velocity dispersion $\sigma_v$ corresponding to a Gaussian line profile, for which $\Delta$v is the full width at half maximum (FWHM). $\mu$ is the mean molecular weight, which we take as 2.8, as stated above \citep{2004ApJ...600..279C,2008A&A...487..993K},
whereas $\mu_0$ is the magnetic permeability of free space.

\begin{table}
\caption{Virial energy balance of the `Spire' and `SFO30' M16 regions.
For equilibrium, the sum of the top three rows
(the left-hand side, LHS, of equation~3) should 
roughly equal the sum of the lower two rows (the right-hand side, RHS, of equation~3).
}
\centering
\begin{tabular}{
    l
    r@{\hspace{0.15em}--\hspace{0.15em}}l
    @{\hspace{1.0em}}
    r@{\hspace{0.15em}--\hspace{0.15em}}l
}
\hline
Parameter
& \multicolumn{2}{c}{Spire / $10^{37}\,\mathrm{J}$}
& \multicolumn{2}{c}{SFO30 / $10^{37}\,\mathrm{J}$} \\
\hline
$2E_{\mathrm{T}}$
& $\sim 1.1$ & $1.5$
& $\sim 1.4$ & $2.1$ \\

$2E_{\mathrm{NT}}$
& $\sim 2.1$ & $4.7$
& $\sim 2.6$ & $5.9$ \\

$E_{\mathrm{B}}$
& $\sim 0.5$ & $1.2$
& $\sim 1.8$ & $4.1$ \\

$E_{\mathrm{G}}$
& $\sim 5.4$ & $7.2$
& $\sim 8.4$ & $10.3$ \\

$E_{\mathrm{X}}$
& $\sim 0.1$ & $0.3$
& $\sim 0.1$ & $0.3$ \\

LHS
& $\sim 3.7$ & $7.4$
& $\sim 5.8$ & $12.1$ \\

RHS
& $\sim 5.5$ & $7.5$
& $\sim 8.5$ & $10.6$ \\
\hline
\end{tabular}
\end{table}

The two regions that we labelled `base' are clearly just parts of the attached pillars that have been well studied previously.
Therefore, for the remainder of this paper we concentrate on the regions we refer to as the `Spire' and `SFO30'. For both of these regions we carry out the
calculations listed in equations 4--8 and list the values
that we obtain in Table~2. 

\subsection{The Spire}

We first concentrate on the region referred to as the `Spire'.
We note that, if we substitute the values from Table 2 into 
equation~3, we obtain a range of values for the left-hand side (LHS)
of $\sim$~3.7 $-$ 7.4, while the right-hand side (RHS) has a range of values of
$\sim$~5.5 $-$ 7.5.
These two ranges are essentially the same (given the size of the uncertainties),
as per the requirement for equilibrium, with the RHS simply having a narrower range, due to the range of uncertainties in the input values.
It is notable that all energies have the same order of magnitude.

This also leads us to the hypothesis
that this may be a bound core and hence could be termed `pre-stellar',
rather than merely `starless' \citep{1994MNRAS.268..276W}.
However, we also note that the contribution of $E_{\mathrm{B}}$
to the `supporting' potential energies, is
in the range $\sim$~15 $-$ 20~per cent, which we claim is significant.

Consequently, we propose that the tip of the Spire contains a pre-stellar core in
approximate virial equilibrium, and that the magnetic field is playing
a significant role in maintaining that equilibrium. This is
similar to the findings of \citet{2018ApJ...860L...6P} in the `Pillars
of Creation' $-$ particularly at the head of `Pillar~2'.

\subsection{SFO30}

Now turning to `SFO30', we note that the LHS of equation~3 in this case is $\sim$~5.8 $-$ 12.1, while the right-hand side is
$\sim$~8.5 $-$ 10.6. Hence, the range of potential energies of the forces trying 
to collapse the core lies entirely within the range of those trying to 
support the core.

We interpret this as another core that is probably bound, but lying in
approximate equilibrium. Once again the magnetic field appears to play a
significant role, contributing $\sim$~30~per cent of the LHS. Hence, it is
being supported significantly by magnetic pressure
as well as gas pressure, against its own self-gravity and the external
pressure of nearby O-type stars, whilst remaining bound.

\section{Summary}

We have mapped the wider region of the Eagle Nebula, M16, in 850-$\mu$m
polarized dust emission, as part of the JCMT BISTRO survey, using the SCUBA-2/POL-2 polarimeter-camera combination. We covered not only the region of the `Pillars of Creation', but also the wider nebula, including in 
particular regions known as the `Spire' and the bright-rimmed cloud
that we refer to as `SFO30'.

We note that both the `Spire' and `SFO30' display a magnetic field pattern that mimics the field pattern seen in the `Pillars'. 
Namely, the field lies along the length of the pillars, and curves around at their tips. `SFO30' shows both the longitudinal field and the curving round at the tip. Whereas the `Spire' only shows the region where the field at the tip lies orthogonal to the direction to the O-type star 
that is causing the pillar-like structure, just like `Pillar~1'.

We find agreement with the formation model of the pillars proposed by \citet{2018ApJ...860L...6P}, in which a shock from a nearby O-type star erodes a cloud around a pre-existing over-density in the cloud and
creates a pillar with a dense core at its tip. During this process, the pre-existing magnetic field is forced to lie along the length of the pillar, curving round at its tip. Both the `Spire' and `SFO30' demonstrate this behaviour.

\rev{
The POL-2 magnetic-field orientations were compared with the larger-scale magnetic-field direction inferred from Planck 353-GHz Stokes $Q$ and $U$ data, as shown in Figure 2. This suggests that the dense-gas field is locally reoriented rather than simply tracing the surrounding Galactic-scale field.
}

We used the Davis-Chandrasekhar-Fermi (DCF) method to estimate the magnetic field strength in these regions, based on the angular dispersion of the polarization vectors. We estimated the magnetic potential energy in each region, and compared it to the pressure energy of the gas (both thermal and non-thermal), and the potential energy of each core's self-gravity, as well as the external pressure exerted by nearby
O-type stars.

Based on this analysis, we found that both the `Spire' and `SFO30' are in approximate equilibrium. In both cases, the magnetic field contributes significantly to the support of the cores, just as was previously found in the pillars.

Consequently, we supported the previous conclusion that the pillars are magnetically-supported and postulated that the previous model for their development is correct. Furthermore, we showed that the pillars are not unique, and here suggest that this might be a fairly common star formation mechanism. Hence, we propose that cores formed in this way are longer-lived than they would otherwise have been, courtesy of the support of their internal magnetic fields.

\section*{Acknowledgements}

D.W.-T. and J.M.K acknowledge Science and Technology Facilities Council (STFC) support under grant number ST/R000786/1. K.P. is a Royal Society University Research Fellow, supported by grant No. URF/R1/211322. J.K. is currently supported by the Royal Society under grant number RF\textbackslash ERE\textbackslash231132, as part of project URF\textbackslash R1\textbackslash211322. The James Clerk Maxwell Telescope is operated by the East Asian Observatory on behalf of the Academia Sinica Institute of Astronomy and Astrophysics, and the National Astronomical Research Institute of Thailand. Additional funding support is provided by the Science and Technology Facilities Council of the United Kingdom and participating universities in the United Kingdom and Canada. The authors wish to recognize and acknowledge the very significant cultural role and reverence that the summit of Maunakea has always had within the indigenous Hawaiian community.  We are most fortunate to have the opportunity to conduct observations from this mountain.

SCUBA-2 and POL-2 were built 
at the UK Astronomy Technology Centre,
partially through grants from the Canada Foundation for Innovation.  
This research used the facilities of the Canadian Astronomy Data Centre operated by the National 
Research Council of Canada with the support of the Canadian Space Agency.   
This research has also made use of NASA's Astrophysics Data System Bibliographic Services, and distances were obtained using data from the European Space Agency (ESA) mission
{\it Gaia} (\url{https://www.cosmos.esa.int/gaia}), processed by the {\it Gaia}
Data Processing and Analysis Consortium (DPAC,
\url{https://www.cosmos.esa.int/web/gaia/dpac/consortium}). Funding for the DPAC
has been provided by national institutions, in particular the institutions
participating in the {\it Gaia} Multilateral Agreement.

\section*{Data Availability}
 
All data are available through the Canadian Astrophysics Data Centre (CADC).

\bibliographystyle{mnras}
\bibliography{M16.bib} 

@article{1951ApJ...114..206D,
	adsurl = {https://ui.adsabs.harvard.edu/abs/1951ApJ...114..206D},
	author = {{Davis}, Jr., Leverett and {Greenstein}, Jesse L.},
	doi = {10.1086/145464},
	journal = {\apj},
	month = sep,
	pages = {206},
	title = {{The Polarization of Starlight by Aligned Dust Grains.}},
	volume = {114},
	year = 1951}

@article{1976Ap&SS..43..291D,
	adsurl = {https://ui.adsabs.harvard.edu/abs/1976Ap&SS..43..291D},
	author = {{Dolginov}, A.~Z. and {Mitrofanov}, I.~G.},
	doi = {10.1007/BF00640010},
	journal = {\apss},
	month = sep,
	number = {2},
	pages = {291-317},
	title = {{Orientation of Cosmic Dust Grains}},
	volume = {43},
	year = 1976}

@article{1949Sci...109..165H,
	adsurl = {https://ui.adsabs.harvard.edu/abs/1949Sci...109..165H},
	author = {{Hiltner}, W.~A.},
	doi = {10.1126/science.109.2825.165},
	journal = {Science},
	month = feb,
	number = {2825},
	pages = {165},
	title = {{Polarization of Light from Distant Stars by Interstellar Medium}},
	volume = {109},
	year = 1949}

@article{1949AJ.....54..187H,
	adsurl = {https://ui.adsabs.harvard.edu/abs/1949AJ.....54..187H},
	author = {{Hall}, John S. and {Mikesell}, Alfred H.},
	doi = {10.1086/106256},
	journal = {\aj},
	month = sep,
	pages = {187-188},
	title = {{Observations of polarized light from stars.}},
	volume = {54},
	year = 1949}

@article{2023AJ....165..198H,
	adsurl = {https://ui.adsabs.harvard.edu/abs/2023AJ....165..198H},
	archiveprefix = {arXiv},
	author = {{Hwang}, Jihye and {Pattle}, Kate and {Parsons}, Harriet and {Go}, Mallory and {Kim}, Jongsoo},
	doi = {10.3847/1538-3881/acc460},
	eid = {198},
	eprint = {2303.07628},
	journal = {\aj},
	month = may,
	number = {5},
	pages = {198},
	primaryclass = {astro-ph.GA},
	title = {{Magnetic Fields in the Horsehead Nebula}},
	volume = {165},
	year = 2023}

@ARTICLE{1991ApJS...77...59S,
       author = {{Sugitani}, Koji and {Fukui}, Yasuo and {Ogura}, Katsuo},
        title = "{A Catalog of Bright-rimmed Clouds with IRAS Point Sources: Candidates for Star Formation by Radiation-driven Implosion. I. The Northern Hemisphere}",
      journal = {\apjs},
         year = 1991,
        month = sep,
       volume = {77},
        pages = {59},
          doi = {10.1086/191597},
       adsurl = {https://ui.adsabs.harvard.edu/abs/1991ApJS...77...59S}
}

@article{2008A&A...487..993K,
	adsurl = {https://ui.adsabs.harvard.edu/abs/2008A&A...487..993K},
	archiveprefix = {arXiv},
	author = {{Kauffmann}, J. and {Bertoldi}, F. and {Bourke}, T.~L. and {Evans}, II, N.~J. and {Lee}, C.~W.},
	doi = {10.1051/0004-6361:200809481},
	eprint = {0805.4205},
	journal = {\aap},
	month = sep,
	number = {3},
	pages = {993-1017},
	primaryclass = {astro-ph},
	title = {{MAMBO mapping of Spitzer c2d small clouds and cores}},
	volume = {487},
	year = 2008}

@article{2019ApJ...870...32K,
	adsurl = {https://ui.adsabs.harvard.edu/abs/2019ApJ...870...32K},
	archiveprefix = {arXiv},
	author = {{Kuhn}, Michael A. and {Hillenbrand}, Lynne A. and {Sills}, Alison and {Feigelson}, Eric D. and {Getman}, Konstantin V.},
	doi = {10.3847/1538-4357/aaef8c},
	eid = {32},
	eprint = {1807.02115},
	journal = {\apj},
	month = jan,
	number = {1},
	pages = {32},
	primaryclass = {astro-ph.GA},
	title = {{Kinematics in Young Star Clusters and Associations with Gaia DR2}},
	volume = {870},
	year = 2019}

@article{2007JQSRT.106..225L,
	adsurl = {https://ui.adsabs.harvard.edu/abs/2007JQSRT.106..225L},
	archiveprefix = {arXiv},
	author = {{Lazarian}, A.},
	doi = {10.1016/j.jqsrt.2007.01.038},
	eprint = {0707.0858},
	journal = {\jqsrt},
	month = jul,
	pages = {225-256},
	primaryclass = {astro-ph},
	title = {{Tracing magnetic fields with aligned grains}},
	volume = {106},
	year = 2007}

@article{2008ApJ...676L..25L,
	adsurl = {https://ui.adsabs.harvard.edu/abs/2008ApJ...676L..25L},
	archiveprefix = {arXiv},
	author = {{Lazarian}, A. and {Hoang}, Thiem},
	doi = {10.1086/586706},
	eprint = {0801.0265},
	journal = {\apjl},
	month = mar,
	number = {1},
	pages = {L25},
	primaryclass = {astro-ph},
	title = {{Alignment of Dust with Magnetic Inclusions: Radiative Torques and Superparamagnetic Barnett and Nuclear Relaxation}},
	volume = {676},
	year = 2008}

@article{2001ApJ...546..980O,
	adsurl = {https://ui.adsabs.harvard.edu/abs/2001ApJ...546..980O},
	archiveprefix = {arXiv},
	author = {{Ostriker}, Eve C. and {Stone}, James M. and {Gammie}, Charles F.},
	doi = {10.1086/318290},
	eprint = {astro-ph/0008454},
	journal = {\apj},
	month = jan,
	number = {2},
	pages = {980-1005},
	primaryclass = {astro-ph},
	title = {{Density, Velocity, and Magnetic Field Structure in Turbulent Molecular Cloud Models}},
	volume = {546},
	year = 2001}

@article{2021ApJ...907...88P,
	adsurl = {https://ui.adsabs.harvard.edu/abs/2021ApJ...907...88P},
	archiveprefix = {arXiv},
	author = {{Pattle}, Kate and {Lai}, Shih-Ping and {Di Francesco}, James and {Sadavoy}, Sarah and {Ward-Thompson}, Derek and {Johnstone}, Doug and {Hoang}, Thiem and {Arzoumanian}, Doris and {Bastien}, Pierre and {Bourke}, Tyler L. and {Coud{\'e}}, Simon and {Doi}, Yasuo and {Eswaraiah}, Chakali and {Fanciullo}, Lapo and {Furuya}, Ray S. and {Hwang}, Jihye and {Hull}, Charles L.~H. and {Kang}, Jihyun and {Kim}, Kee-Tae and {Kirchschlager}, Florian and {Kwon}, Jungmi and {Kwon}, Woojin and {Lee}, Chang Won and {Liu}, Tie and {Redman}, Matt and {Soam}, Archana and {Tahani}, Mehrnoosh and {Tamura}, Motohide and {Tang}, Xindi},
	doi = {10.3847/1538-4357/abcc6c},
	eid = {88},
	eprint = {2011.09765},
	journal = {\apj},
	month = feb,
	number = {2},
	pages = {88},
	primaryclass = {astro-ph.GA},
	title = {{JCMT POL-2 and BISTRO Survey Observations of Magnetic Fields in the L1689 Molecular Cloud}},
	volume = {907},
	year = 2021}

@inproceedings{2023ASPC..534..193P,
	adsurl = {https://ui.adsabs.harvard.edu/abs/2023ASPC..534..193P},
	archiveprefix = {arXiv},
	author = {{Pattle}, K. and {Fissel}, L. and {Tahani}, M. and {Liu}, T. and {Ntormousi}, E.},
	booktitle = {Protostars and Planets VII},
	doi = {10.48550/arXiv.2203.11179},
	editor = {{Inutsuka}, S. and {Aikawa}, Y. and {Muto}, T. and {Tomida}, K. and {Tamura}, M.},
	eprint = {2203.11179},
	month = jul,
	pages = {193},
	primaryclass = {astro-ph.GA},
	series = {Astronomical Society of the Pacific Conference Series},
	title = {{Magnetic Fields in Star Formation: from Clouds to Cores}},
	volume = {534},
	year = 2023}

@article{1997A&A...322..598S,
	adsurl = {https://ui.adsabs.harvard.edu/abs/1997A&A...322..598S},
	archiveprefix = {arXiv},
	author = {{Schaerer}, D. and {de Koter}, A.},
	doi = {10.48550/arXiv.astro-ph/9611068},
	eprint = {astro-ph/9611068},
	journal = {\aap},
	month = jun,
	pages = {598-614},
	primaryclass = {astro-ph},
	title = {{Combined stellar structure and atmosphere models for massive stars. III. Spectral evolution and revised ionizing fluxes of O3-B0 stars.}},
	volume = {322},
	year = 1997}

@article{1954ApJ...120....1S,
	adsurl = {https://ui.adsabs.harvard.edu/abs/1954ApJ...120....1S},
	author = {{Spitzer}, Jr., Lyman},
	doi = {10.1086/145876},
	journal = {\apj},
	month = jul,
	pages = {1},
	title = {{Behavior of Matter in Space.}},
	volume = {120},
	year = 1954}

@article{2006MNRAS.369.1201W,
	adsurl = {https://ui.adsabs.harvard.edu/abs/2006MNRAS.369.1201W},
	archiveprefix = {arXiv},
	author = {{Ward-Thompson}, D. and {Nutter}, D. and {Bontemps}, S. and {Whitworth}, A. and {Attwood}, R.},
	doi = {10.1111/j.1365-2966.2006.10356.x},
	eprint = {astro-ph/0603604},
	journal = {\mnras},
	month = jul,
	number = {3},
	pages = {1201-1210},
	primaryclass = {astro-ph},
	title = {{SCUBA observations of the Horsehead nebula - what did the horse swallow?}},
	volume = {369},
	year = 2006}

@book{2011isf..book.....W,
	adsurl = {https://ui.adsabs.harvard.edu/abs/2011isf..book.....W},
	author = {{Ward-Thompson}, Derek and {Whitworth}, Anthony P.},
	title = {{An Introduction to Star Formation}},
	year = 2011}

@article{1999A&A...342..233W,
	adsurl = {https://ui.adsabs.harvard.edu/abs/1999A&A...342..233W},
	author = {{White}, G.~J. and {Nelson}, R.~P. and {Holland}, W.~S. and {Robson}, E.~I. and {Greaves}, J.~S. and {McCaughrean}, M.~J. and {Pilbratt}, G.~L. and {Balser}, D.~S. and {Oka}, T. and {Sakamoto}, S. and {Hasegawa}, T. and {McCutcheon}, W.~H. and {Matthews}, H.~E. and {Fridlund}, C.~V.~M. and {Tothill}, N.~F.~H. and {Huldtgren}, M. and {Deane}, J.~R.},
	journal = {\aap},
	month = feb,
	pages = {233-256},
	title = {{The Eagle Nebula's fingers - pointers to the earliest stages of star formation?}},
	volume = {342},
	year = 1999}

@article{2001MNRAS.327..788W,
	adsurl = {https://ui.adsabs.harvard.edu/abs/2001MNRAS.327..788W},
	archiveprefix = {arXiv},
	author = {{Williams}, R.~J.~R. and {Ward-Thompson}, D. and {Whitworth}, A.~P.},
	doi = {10.1046/j.1365-8711.2001.04757.x},
	eprint = {astro-ph/0107272},
	journal = {\mnras},
	month = nov,
	number = {3},
	pages = {788-798},
	primaryclass = {astro-ph},
	title = {{Hydrodynamics of photoionized columns in the Eagle Nebula, M 16}},
	volume = {327},
	year = 2001}

@article{2007ARA&A..45..481Z,
	adsurl = {https://ui.adsabs.harvard.edu/abs/2007ARA&A..45..481Z},
	archiveprefix = {arXiv},
	author = {{Zinnecker}, Hans and {Yorke}, Harold W.},
	doi = {10.1146/annurev.astro.44.051905.092549},
	eprint = {0707.1279},
	journal = {\araa},
	month = sep,
	number = {1},
	pages = {481-563},
	primaryclass = {astro-ph},
	title = {{Toward Understanding Massive Star Formation}},
	volume = {45},
	year = 2007}

@article{1953ApJ...118..113C,
	adsurl = {https://ui.adsabs.harvard.edu/abs/1953ApJ...118..113C},
	author = {{Chandrasekhar}, S. and {Fermi}, E.},
	doi = {10.1086/145731},
	journal = {\apj},
	month = jul,
	pages = {113},
	title = {{Magnetic Fields in Spiral Arms.}},
	volume = {118},
	year = 1953}

@article{2012ARA&A..50...29C,
	adsurl = {https://ui.adsabs.harvard.edu/abs/2012ARA&A..50...29C},
	author = {{Crutcher}, Richard M.},
	doi = {10.1146/annurev-astro-081811-125514},
	journal = {\araa},
	month = sep,
	pages = {29-63},
	title = {{Magnetic Fields in Molecular Clouds}},
	volume = {50},
	year = 2012}

@article{2004ApJ...600..279C,
	adsurl = {https://ui.adsabs.harvard.edu/abs/2004ApJ...600..279C},
	archiveprefix = {arXiv},
	author = {{Crutcher}, Richard M. and {Nutter}, D.~J. and {Ward-Thompson}, D. and {Kirk}, J.~M.},
	doi = {10.1086/379705},
	eprint = {astro-ph/0305604},
	journal = {\apj},
	month = jan,
	number = {1},
	pages = {279-285},
	primaryclass = {astro-ph},
	title = {{SCUBA Polarization Measurements of the Magnetic Field Strengths in the L183, L1544, and L43 Prestellar Cores}},
	volume = {600},
	year = 2004}

@article{1951PhRv...81..890D,
	adsurl = {https://ui.adsabs.harvard.edu/abs/1951PhRv...81..890D},
	author = {{Davis}, Leverett},
	doi = {10.1103/PhysRev.81.890.2},
	journal = {Physical Review},
	month = mar,
	number = {5},
	pages = {890-891},
	title = {{The Strength of Interstellar Magnetic Fields}},
	volume = {81},
	year = 1951}

@article{2017ApJ...842...66W,
	adsurl = {https://ui.adsabs.harvard.edu/abs/2017ApJ...842...66W},
	archiveprefix = {arXiv},
	author = {{Ward-Thompson}, Derek and {Pattle}, Kate and {Bastien}, Pierre and {Furuya}, Ray S. and {Kwon}, Woojin and {Lai}, Shih-Ping and {Qiu}, Keping and {Berry}, David and {Choi}, Minho and {Coud{\'e}}, Simon and {Di Francesco}, James and {Hoang}, Thiem and {Franzmann}, Erica and {Friberg}, Per and {Graves}, Sarah F. and {Greaves}, Jane S. and {Houde}, Martin and {Johnstone}, Doug and {Kirk}, Jason M. and {Koch}, Patrick M. and {Kwon}, Jungmi and {Lee}, Chang Won and {Li}, Di and {Matthews}, Brenda C. and {Mottram}, Joseph C. and {Parsons}, Harriet and {Pon}, Andy and {Rao}, Ramprasad and {Rawlings}, Mark and {Shinnaga}, Hiroko and {Sadavoy}, Sarah and {van Loo}, Sven and {Aso}, Yusuke and {Byun}, Do-Young and {Eswaraiah}, Chakali and {Chen}, Huei-Ru and {Chen}, Mike C. -Y. and {Chen}, Wen Ping and {Ching}, Tao-Chung and {Cho}, Jungyeon and {Chrysostomou}, Antonio and {Chung}, Eun Jung and {Doi}, Yasuo and {Drabek-Maunder}, Emily and {Eyres}, Stewart P.~S. and {Fiege}, Jason and {Friesen}, Rachel K. and {Fuller}, Gary and {Gledhill}, Tim and {Griffin}, Matt J. and {Gu}, Qilao and {Hasegawa}, Tetsuo and {Hatchell}, Jennifer and {Hayashi}, Saeko S. and {Holland}, Wayne and {Inoue}, Tsuyoshi and {Inutsuka}, Shu-ichiro and {Iwasaki}, Kazunari and {Jeong}, Il-Gyo and {Kang}, Ji-hyun and {Kang}, Miju and {Kang}, Sung-ju and {Kawabata}, Koji S. and {Kemper}, Francisca and {Kim}, Gwanjeong and {Kim}, Jongsoo and {Kim}, Kee-Tae and {Kim}, Kyoung Hee and {Kim}, Mi-Ryang and {Kim}, Shinyoung and {Lacaille}, Kevin M. and {Lee}, Jeong-Eun and {Lee}, Sang-Sung and {Li}, Dalei and {Li}, Hua-bai and {Liu}, Hong-Li and {Liu}, Junhao and {Liu}, Sheng-Yuan and {Liu}, Tie and {Lyo}, A. -Ran and {Mairs}, Steve and {Matsumura}, Masafumi and {Moriarty-Schieven}, Gerald H. and {Nakamura}, Fumitaka and {Nakanishi}, Hiroyuki and {Ohashi}, Nagayoshi and {Onaka}, Takashi and {Peretto}, Nicolas and {Pyo}, Tae-Soo and {Qian}, Lei and {Retter}, Brendan and {Richer}, John and {Rigby}, Andrew and {Robitaille}, Jean-Fran{\c{c}}ois and {Savini}, Giorgio and {Scaife}, Anna M.~M. and {Soam}, Archana and {Tamura}, Motohide and {Tang}, Ya-Wen and {Tomisaka}, Kohji and {Wang}, Hongchi and {Wang}, Jia-Wei and {Whitworth}, Anthony P. and {Yen}, Hsi-Wei and {Yoo}, Hyunju and {Yuan}, Jinghua and {Zhang}, Chuan-Peng and {Zhang}, Guoyin and {Zhou}, Jianjun and {Zhu}, Lei and {Andr{\'e}}, Philippe and {Dowell}, C. Darren and {Falle}, Sam and {Tsukamoto}, Yusuke},
	doi = {10.3847/1538-4357/aa70a0},
	eid = {66},
	eprint = {1704.08552},
	journal = {\apj},
	month = jun,
	number = {1},
	pages = {66},
	primaryclass = {astro-ph.GA},
	title = {{First Results from BISTRO: A SCUBA-2 Polarimeter Survey of the Gould Belt}},
	volume = {842},
	year = 2017}

@article{2018ApJ...860L...6P,
	adsurl = {https://ui.adsabs.harvard.edu/abs/2018ApJ...860L...6P},
	archiveprefix = {arXiv},
	author = {{Pattle}, Kate and {Ward-Thompson}, Derek and {Hasegawa}, Tetsuo and {Bastien}, Pierre and {Kwon}, Woojin and {Lai}, Shih-Ping and {Qiu}, Keping and {Furuya}, Ray and {Berry}, David and {JCMT BISTRO Survey Team}},
	doi = {10.3847/2041-8213/aac771},
	eid = {L6},
	eprint = {1805.11554},
	journal = {\apjl},
	month = jun,
	number = {1},
	pages = {L6},
	primaryclass = {astro-ph.GA},
	title = {{First Observations of the Magnetic Field inside the Pillars of Creation: Results from the BISTRO Survey}},
	volume = {860},
	year = 2018}

@INPROCEEDINGS{2016SPIE.9914E..03F,
       author = {{Friberg}, Per and {Bastien}, Pierre and {Berry}, David and {Savini}, Giorgio and {Graves}, Sarah F. and {Pattle}, Kate},
        title = "{POL-2: a polarimeter for the James-Clerk-Maxwell telescope}",
    booktitle = {Millimeter, Submillimeter, and Far-Infrared Detectors and Instrumentation for Astronomy VIII},
         year = 2016,
       editor = {{Holland}, Wayne S. and {Zmuidzinas}, Jonas},
       series = {Society of Photo-Optical Instrumentation Engineers (SPIE) Conference Series},
       volume = {9914},
        month = jul,
          eid = {991403},
        pages = {991403},
          doi = {10.1117/12.2231943},
       adsurl = {https://ui.adsabs.harvard.edu/abs/2016SPIE.9914E..03F}
}

@ARTICLE{2022MNRAS.514.1575C,
       author = {{Chen}, Che-Yu and {Li}, Zhi-Yun and {Mazzei}, Renato R. and {Park}, Jinsoo and {Fissel}, Laura M. and {Chen}, Michael C.-Y. and {Klein}, Richard I. and {Li}, Pak Shing},
        title = "{The Davis-Chandrasekhar-Fermi method revisited}",
      journal = {\mnras},
         year = 2022,
        month = aug,
       volume = {514},
       number = {2},
        pages = {1575-1594},
          doi = {10.1093/mnras/stac1417},
archivePrefix = {arXiv},
       eprint = {2205.09134},
 primaryClass = {astro-ph.GA},
       adsurl = {https://ui.adsabs.harvard.edu/abs/2022MNRAS.514.1575C}
}

@ARTICLE{1994MNRAS.268..276W,
       author = {{Ward-Thompson}, D. and {Scott}, P.~F. and {Hills}, R.~E. and {Andre}, P.},
        title = "{A Submillimetre Continuum Survey of Pre Protostellar Cores}",
      journal = {\mnras},
         year = 1994,
        month = may,
       volume = {268},
        pages = {276},
          doi = {10.1093/mnras/268.1.276},
       adsurl = {https://ui.adsabs.harvard.edu/abs/1994MNRAS.268..276W}
}

@article{Friesen_2017,
doi = {10.3847/1538-4357/aa6d58},
url = {https://doi.org/10.3847/1538-4357/aa6d58},
year = {2017},
month = {jul},
publisher = {The American Astronomical Society},
volume = {843},
number = {1},
pages = {63},
author = {Friesen, Rachel K. and Pineda, Jaime E. and (co-PIs) and Rosolowsky, Erik and Alves, Felipe and Chacón-Tanarro, Ana and Chen, Hope How-Huan and Chen, Michael Chun-Yuan and Di Francesco, James and Keown, Jared and Kirk, Helen and Punanova, Anna and Seo, Youngmin and Shirley, Yancy and Ginsburg, Adam and Hall, Christine and Offner, Stella S. R. and Singh, Ayushi and Arce, Héctor G. and Caselli, Paola and Goodman, Alyssa A. and Martin, Peter G. and Matzner, Christopher and Myers, Philip C. and Redaelli, Elena and (The GAS Collaboration)},
title = {The Green Bank Ammonia Survey: First Results of NH3 Mapping of the Gould Belt},
journal = {The Astrophysical Journal}
}

@ARTICLE{2001ApJ...547..317W,
       author = {{Whitworth}, A.~P. and {Ward-Thompson}, D.},
        title = "{An Empirical Model for Protostellar Collapse}",
      journal = {\apj},
         year = 2001,
        month = jan,
       volume = {547},
       number = {1},
        pages = {317-322},
          doi = {10.1086/318373},
archivePrefix = {arXiv},
       eprint = {astro-ph/0009325},
 primaryClass = {astro-ph},
       adsurl = {https://ui.adsabs.harvard.edu/abs/2001ApJ...547..317W}
}

@article{Telford_2023,
doi = {10.3847/1538-4357/aca896},
url = {https://doi.org/10.3847/1538-4357/aca896},
year = {2023},
month = {jan},
publisher = {The American Astronomical Society},
volume = {943},
number = {1},
pages = {65},
author = {Telford, O. Grace and McQuinn, Kristen B. W. and Chisholm, John and Berg, Danielle A.},
title = {The Ionizing Spectra of Extremely Metal-poor O Stars: Constraints from the Only H ii Region in Leo P},
journal = {The Astrophysical Journal}
}

@ARTICLE{2016A&A...595A...1G,
       author = {{Gaia Collaboration} and {Prusti}, T. and {de Bruijne}, J.~H.~J. and {Brown}, A.~G.~A. and {Vallenari}, A. and {Babusiaux}, C. and {Bailer-Jones}, C.~A.~L. and {Bastian}, U. and {Biermann}, M. and {Evans}, D.~W. and {Eyer}, L. and {Jansen}, F. and {Jordi}, C. and {Klioner}, S.~A. and {Lammers}, U. and {Lindegren}, L. and {Luri}, X. and {Mignard}, F. and {Milligan}, D.~J. and {Panem}, C. and {Poinsignon}, V. and {Pourbaix}, D. and {Randich}, S. and {Sarri}, G. and {Sartoretti}, P. and {Siddiqui}, H.~I. and {Soubiran}, C. and {Valette}, V. and {van Leeuwen}, F. and {Walton}, N.~A. and {Aerts}, C. and {Arenou}, F. and {Cropper}, M. and {Drimmel}, R. and {H{\o}g}, E. and {Katz}, D. and {Lattanzi}, M.~G. and {O'Mullane}, W. and {Grebel}, E.~K. and {Holland}, A.~D. and {Huc}, C. and {Passot}, X. and {Bramante}, L. and {Cacciari}, C. and {Casta{\~n}eda}, J. and {Chaoul}, L. and {Cheek}, N. and {De Angeli}, F. and {Fabricius}, C. and {Guerra}, R. and {Hern{\'a}ndez}, J. and {Jean-Antoine-Piccolo}, A. and {Masana}, E. and {Messineo}, R. and {Mowlavi}, N. and {Nienartowicz}, K. and {Ord{\'o}{\~n}ez-Blanco}, D. and {Panuzzo}, P. and {Portell}, J. and {Richards}, P.~J. and {Riello}, M. and {Seabroke}, G.~M. and {Tanga}, P. and {Th{\'e}venin}, F. and {Torra}, J. and {Els}, S.~G. and {Gracia-Abril}, G. and {Comoretto}, G. and {Garcia-Reinaldos}, M. and {Lock}, T. and {Mercier}, E. and {Altmann}, M. and {Andrae}, R. and {Astraatmadja}, T.~L. and {Bellas-Velidis}, I. and {Benson}, K. and {Berthier}, J. and {Blomme}, R. and {Busso}, G. and {Carry}, B. and {Cellino}, A. and {Clementini}, G. and {Cowell}, S. and {Creevey}, O. and {Cuypers}, J. and {Davidson}, M. and {De Ridder}, J. and {de Torres}, A. and {Delchambre}, L. and {Dell'Oro}, A. and {Ducourant}, C. and {Fr{\'e}mat}, Y. and {Garc{\'\i}a-Torres}, M. and {Gosset}, E. and {Halbwachs}, J.-L. and {Hambly}, N.~C. and {Harrison}, D.~L. and {Hauser}, M. and {Hestroffer}, D. and {Hodgkin}, S.~T. and {Huckle}, H.~E. and {Hutton}, A. and {Jasniewicz}, G. and {Jordan}, S. and {Kontizas}, M. and {Korn}, A.~J. and {Lanzafame}, A.~C. and {Manteiga}, M. and {Moitinho}, A. and {Muinonen}, K. and {Osinde}, J. and {Pancino}, E. and {Pauwels}, T. and {Petit}, J.-M. and {Recio-Blanco}, A. and {Robin}, A.~C. and {Sarro}, L.~M. and {Siopis}, C. and {Smith}, M. and {Smith}, K.~W. and {Sozzetti}, A. and {Thuillot}, W. and {van Reeven}, W. and {Viala}, Y. and {Abbas}, U. and {Abreu Aramburu}, A. and {Accart}, S. and {Aguado}, J.~J. and {Allan}, P.~M. and {Allasia}, W. and {Altavilla}, G. and {{\'A}lvarez}, M.~A. and {Alves}, J. and {Anderson}, R.~I. and {Andrei}, A.~H. and {Anglada Varela}, E. and {Antiche}, E. and {Antoja}, T. and {Ant{\'o}n}, S. and {Arcay}, B. and {Atzei}, A. and {Ayache}, L. and {Bach}, N. and {Baker}, S.~G. and {Balaguer-N{\'u}{\~n}ez}, L. and {Barache}, C. and {Barata}, C. and {Barbier}, A. and {Barblan}, F. and {Baroni}, M. and {Barrado y Navascu{\'e}s}, D. and {Barros}, M. and {Barstow}, M.~A. and {Becciani}, U. and {Bellazzini}, M. and {Bellei}, G. and {Bello Garc{\'\i}a}, A. and {Belokurov}, V. and {Bendjoya}, P. and {Berihuete}, A. and {Bianchi}, L. and {Bienaym{\'e}}, O. and {Billebaud}, F. and {Blagorodnova}, N. and {Blanco-Cuaresma}, S. and {Boch}, T. and {Bombrun}, A. and {Borrachero}, R. and {Bouquillon}, S. and {Bourda}, G. and {Bouy}, H. and {Bragaglia}, A. and {Breddels}, M.~A. and {Brouillet}, N. and {Br{\"u}semeister}, T. and {Bucciarelli}, B. and {Budnik}, F. and {Burgess}, P. and {Burgon}, R. and {Burlacu}, A. and {Busonero}, D. and {Buzzi}, R. and {Caffau}, E. and {Cambras}, J. and {Campbell}, H. and {Cancelliere}, R. and {Cantat-Gaudin}, T. and {Carlucci}, T. and {Carrasco}, J.~M. and {Castellani}, M. and {Charlot}, P. and {Charnas}, J. and {Charvet}, P. and {Chassat}, F. and {Chiavassa}, A. and {Clotet}, M. and {Cocozza}, G. and {Collins}, R.~S. and {Collins}, P. and {Costigan}, G.},
        title = "{The Gaia mission}",
      journal = {\aap},
         year = 2016,
        month = nov,
       volume = {595},
          eid = {A1},
        pages = {A1},
          doi = {10.1051/0004-6361/201629272},
archivePrefix = {arXiv},
       eprint = {1609.04153},
 primaryClass = {astro-ph.IM},
       adsurl = {https://ui.adsabs.harvard.edu/abs/2016A&A...595A...1G}
}

@ARTICLE{2013MNRAS.430.2513H,
       author = {{Holland}, W.~S. and {Bintley}, D. and {Chapin}, E.~L. and {Chrysostomou}, A. and {Davis}, G.~R. and {Dempsey}, J.~T. and {Duncan}, W.~D. and {Fich}, M. and {Friberg}, P. and {Halpern}, M. and {Irwin}, K.~D. and {Jenness}, T. and {Kelly}, B.~D. and {MacIntosh}, M.~J. and {Robson}, E.~I. and {Scott}, D. and {Ade}, P.~A.~R. and {Atad-Ettedgui}, E. and {Berry}, D.~S. and {Craig}, S.~C. and {Gao}, X. and {Gibb}, A.~G. and {Hilton}, G.~C. and {Hollister}, M.~I. and {Kycia}, J.~B. and {Lunney}, D.~W. and {McGregor}, H. and {Montgomery}, D. and {Parkes}, W. and {Tilanus}, R.~P.~J. and {Ullom}, J.~N. and {Walther}, C.~A. and {Walton}, A.~J. and {Woodcraft}, A.~L. and {Amiri}, M. and {Atkinson}, D. and {Burger}, B. and {Chuter}, T. and {Coulson}, I.~M. and {Doriese}, W.~B. and {Dunare}, C. and {Economou}, F. and {Niemack}, M.~D. and {Parsons}, H.~A.~L. and {Reintsema}, C.~D. and {Sibthorpe}, B. and {Smail}, I. and {Sudiwala}, R. and {Thomas}, H.~S.},
        title = "{SCUBA-2: the 10 000 pixel bolometer camera on the James Clerk Maxwell Telescope}",
      journal = {\mnras},
         year = 2013,
        month = apr,
       volume = {430},
       number = {4},
        pages = {2513-2533},
          doi = {10.1093/mnras/sts612},
archivePrefix = {arXiv},
       eprint = {1301.3650},
 primaryClass = {astro-ph.IM},
       adsurl = {https://ui.adsabs.harvard.edu/abs/2013MNRAS.430.2513H}
}

@ARTICLE{2020ApJ...899...28D,
       author = {{Doi}, Yasuo and {Hasegawa}, Tetsuo and {Furuya}, Ray S. and {Coud{\'e}}, Simon and {Hull}, Charles L.~H. and {Arzoumanian}, Doris and {Bastien}, Pierre and {Chen}, Michael Chun-Yuan and {Di Francesco}, James and {Friesen}, Rachel and {Houde}, Martin and {Inutsuka}, Shu-ichiro and {Mairs}, Steve and {Matsumura}, Masafumi and {Onaka}, Takashi and {Sadavoy}, Sarah and {Shimajiri}, Yoshito and {Tahani}, Mehrnoosh and {Tomisaka}, Kohji and {Eswaraiah}, Chakali and {Koch}, Patrick M. and {Pattle}, Kate and {Won Lee}, Chang and {Tamura}, Motohide and {Berry}, David and {Ching}, Tao-Chung and {Hwang}, Jihye and {Kwon}, Woojin and {Soam}, Archana and {Wang}, Jia-Wei and {Lai}, Shih-Ping and {Qiu}, Keping and {Ward-Thompson}, Derek and {Byun}, Do-Young and {Chen}, Huei-Ru Vivien and {Chen}, Wen Ping and {Chen}, Zhiwei and {Cho}, Jungyeon and {Choi}, Minho and {Choi}, Yunhee and {Chrysostomou}, Antonio and {Chung}, Eun Jung and {Diep}, Pham Ngoc and {Duan}, Hao-Yuan and {Fanciullo}, Lapo and {Fiege}, Jason and {Franzmann}, Erica and {Friberg}, Per and {Fuller}, Gary and {Gledhill}, Tim and {Graves}, Sarah F. and {Greaves}, Jane S. and {Griffin}, Matt J. and {Gu}, Qilao and {Han}, Ilseung and {Hatchell}, Jennifer and {Hayashi}, Saeko S. and {Hoang}, Thiem and {Inoue}, Tsuyoshi and {Iwasaki}, Kazunari and {Jeong}, Il-Gyo and {Johnstone}, Doug and {Kanamori}, Yoshihiro and {Kang}, Ji-hyun and {Kang}, Miju and {Kang}, Sung-ju and {Kataoka}, Akimasa and {Kawabata}, Koji S. and {Kemper}, Francisca and {Kim}, Gwanjeong and {Kim}, Jongsoo and {Kim}, Kee-Tae and {Kim}, Kyoung Hee and {Kim}, Mi-Ryang and {Kim}, Shinyoung and {Kirk}, Jason M. and {Kobayashi}, Masato I.~N. and {Konyves}, Vera and {Kusune}, Takayoshi and {Kwon}, Jungmi and {Lacaille}, Kevin and {Law}, Chi-Yan and {Lee}, Chin-Fei and {Lee}, Hyeseung and {Lee}, Jeong-Eun and {Lee}, Sang-Sung and {Lee}, Yong-Hee and {Li}, Dalei and {Li}, Di and {Li}, Hua-bai and {Liu}, Hong-Li and {Liu}, Junhao and {Liu}, Sheng-Yuan and {Liu}, Tie and {de Looze}, Ilse and {Lyo}, A.-Ran and {Matthews}, Brenda C. and {Moriarty-Schieven}, Gerald H. and {Nagata}, Tetsuya and {Nakamura}, Fumitaka and {Nakanishi}, Hiroyuki and {Ohashi}, Nagayoshi and {Park}, Geumsook and {Parsons}, Harriet and {Peretto}, Nicolas and {Pyo}, Tae-Soo and {Qian}, Lei and {Rao}, Ramprasad and {Rawlings}, Mark G. and {Retter}, Brendan and {Richer}, John and {Rigby}, Andrew and {Saito}, Hiro and {Savini}, Giorgio and {Scaife}, Anna M.~M. and {Seta}, Masumichi and {Shinnaga}, Hiroko and {Tang}, Ya-Wen and {Tsukamoto}, Yusuke and {Viti}, Serena and {Wang}, Hongchi and {Whitworth}, Anthony P. and {Yen}, Hsi-Wei and {Yoo}, Hyunju and {Yuan}, Jinghua and {Yun}, Hyeong-Sik and {Zenko}, Tetsuya and {Zhang}, Chuan-Peng and {Zhang}, Guoyin and {Zhang}, Yapeng and {Zhou}, Jianjun and {Zhu}, Lei and {Andr{\'e}}, Philippe and {Dowell}, C. Darren and {Eyres}, Stewart P.~S. and {Falle}, Sam and {van Loo}, Sven and {Robitaille}, Jean-Fran{\c{c}}ois},
        title = "{The JCMT BISTRO Survey: Magnetic Fields Associated with a Network of Filaments in NGC 1333}",
      journal = {\apj},
         year = 2020,
        month = aug,
       volume = {899},
       number = {1},
          eid = {28},
        pages = {28},
          doi = {10.3847/1538-4357/aba1e2},
archivePrefix = {arXiv},
       eprint = {2007.00176},
 primaryClass = {astro-ph.GA},
       adsurl = {https://ui.adsabs.harvard.edu/abs/2020ApJ...899...28D}
}
\label{lastpage}
\end{document}